\documentclass[10pt]{article}
\usepackage{fullpage,graphicx,subfigure,mathdots,mathpazo,color}
\usepackage{amsmath,amscd,tikz,mathrsfs,cite}
\usepackage[normalem]{ulem}
\usepackage{amsmath}
\usepackage{setspace}

\usepackage{epsfig,amsmath,graphicx,amssymb,overpic}
\usepackage{bm}
\usepackage{graphicx}
\usepackage{subfigure, xcolor}
\usepackage{ntheorem}

\def\be{\begin{equation}}
\def\ee{\end{equation}}
\def\bee{\begin{eqnarray}}
\def\ene{\end{eqnarray}}
\def\bes{\begin{subequations}}
\def\ees{\end{subequations}}
\def\no{\nonumber}

\def\d{\displaystyle}

\def\v{\vspace{0.1in}}
\def\no{{\nonumber}}

\allowdisplaybreaks[4]

\usepackage{diagbox}
\begin{document}
\baselineskip=13pt
\renewcommand {\thefootnote}{\dag}
\renewcommand {\thefootnote}{\ddag}
\renewcommand {\thefootnote}{ }

\pagestyle{plain}

\begin{center}
\baselineskip=16pt \leftline{} \vspace{-.3in} {\Large \bf  Dynamics of fractional $N$-soliton solutions with anomalous dispersions
of integrable fractional higher-order nonlinear Schr\"odinger equations} \\[0.2in]
\end{center}

\begin{center}
Weifang Weng$^{1,2}$,\,\, Minghe Zhang$^{1,2}$,\,\, and Zhenya Yan$^{1,2,*}$\footnote{$^{*}${\it Email address}: zyyan@mmrc.iss.ac.cn}  \\[0.05in]
{\it \small $^1$Key Lab of Mathematics Mechanization, Academy of Mathematics and Systems Science, \\ Chinese Academy of Sciences, Beijing 100190, China \\
$^2$School of Mathematical Sciences, University of Chinese Academy of Sciences, Beijing 100049, China} \\
\end{center}

\vspace{0.1in}

{\baselineskip=14pt


\vspace{-0.08in}

In this paper, we explore the anomalous dispersive relations, inverse scattering transform and fractional $N$-soliton solutions of the integrable fractional higher-order nonlinear Schr\"odinger (fHONLS) equations, containing the fractional Hirota (fHirota), fractional complex mKdV (fcmKdV), and fractional Lakshmanan-Porsezian-Daniel (fLPD) equations, etc. The inverse scattering problem can be solved exactly by means of the matrix Riemann-Hilbert problem with simple poles. As a consequence, an explicit formula is found for the fractional $N$-soliton solutions of the fHONLS equations in the reflectionless case. In particular, we analyze the fractional one-, two- and three-soliton solutions with anomalous dispersions of fHirota and fcmKdV equations. The wave, group, and phase velocities of these envelope fractional 1-soliton solutions are related to the power laws of their amplitudes. These obtained fractional $N$-soliton solutions may be useful to explain the related super-dispersion transports of nonlinear waves in fractional nonlinear media.



\vspace{0.3in}

{\bf Fractional nonlinear equations and integrable (integer-order) nonlinear equations are two kinds of important physical models in the fields of nonlinear dynamics and applications. The formers are used to describe physical phenomena with anomalous diffusion, and in general non-integrable such that they can be solved approximately by numerical methods, however, the latters are a class of important physical models and can be solved exactly by the inverse scattering transform (IST) to generate exact solitons, which can be used to compared with numerical and experimental results. More recently, based on two significant aspects, i.e., Riesz fractional derivative and IST integrability, Ablowitz {\it et al} presented the new types of integrable fractional nonlinear soliton equations such as the fractional KdV, fractional NLS,  fractional mKdV, fractional sine-Gordon, and fractional sinh-Gordon equations. Moreover, their fractional one-soliton solutions were found. These solitons show the anomalous dispersions. In this paper, motivated by the idea, we will investigate the integrable fractional extensions of higher-order NLS (fHONLS) equations, containing the fractional Hirota (fHirota), fractional complex mKdV (fcmKdV), fractional LPD (fLPD) equations, and etc. We give the anomalous dispersive relations, and explicit forms of these fHONLS equations via the completeness of eigenfunctions. Based on the IST with  matrix RH problems, we find a formula of fractional $N$-soliton solutions.  In particular, we analyze the fractional one-, two- and three-soliton solutions with anomalous dispersions of fHirota and fcmKdV equations. The wave, group, and phase velocities of these fractional solitons are related to the power laws of their amplitudes. These obtained fractional $N$-soliton solutions may be useful to explain the super-dispersion transports of nonlinear waves in fractional nonlinear media. }

\section{Introduction}

Since the inverse scattering transform (IST), as a nonlinear extension of Fourier transform, was presented by GGKM~\cite{Gardner1967} in 1967, and then the Lax pairs were coined by Lax in 1968~\cite{lax}, many types of nonlinear wave equations have been shown to be IST integrable (i.e., they can be solved exactly by the IST), containing the integrable nonlinear partial differential, differential-difference, difference, and differential-integrable equations (see, e.g., Refs.~\cite{Faddeev1987,ist1,ist2,ist3} and references therein), where the partial derivatives are usually
integer-order derivatives. In fact, since fractional calculus (FC), as an extension of integer-order one, was coined in the L'Hopital's letter written to Leibniz in 1659, more and more attention has been paid to FC and its application in many physical systems with anomalous diffusion such as quantum mechanics, nanofluids, geotechnical engineering, viscoelastic material, and  polymer Science (see, e.g., Refs.~\cite{fc-book1,fc-book2,fc-book,pr20,prl87,west97} and reference therein). Up to now, there are many types of fractional linear equations such as the fractional Schr\"odinger equation~\cite{fls00,long15,zhong16}, and fractional nonlinear wave (fNLW) equations  such as the fractional nonlinear Schr\"odinger (NLS) equation~\cite{carr18,boris20,boris21}
\bee
 i\psi_t+(-\nabla^2)^{\epsilon}\psi+V({\bf r},t)\psi+f(|\psi|^2)\psi=0,\quad {\bf r}\in \mathbb{R}^d,\,\, \epsilon\in (0, 1),
\ene
 where $\nabla^2$ denotes the $d$-dimensional Laplace operator, and the Riesz fractional derivative is defined by~\cite{li19,Riesz,Riesz2}
 \bee \no
 (-\nabla^2)^{\epsilon}\psi({\bf r},t)=\frac{1}{2\pi}\int_{\mathbb{R}^d}d{\bf k}|{\bf k}|^{2\epsilon}\int_{\mathbb{R}^d}e^{i{\bf k}({\bf r}-{\bf r}')}\psi({\bf r}')d{\bf r}',
 \ene
and the fractional complex Ginzburg–Landau equation, etc~\cite{boris21}. However, these fNLW equations were not integrable in the sense of IST such that their exact solutions can usually not found, and their approximate solutions were given with the aid of numerical methods.

More recently, Ablowitz {\it et al}~\cite{ab-prl22,ab22} extended the well-established Riesz fractional derivative~\cite{Riesz,Riesz2,li19} (e.g., $|-\partial^2|^{\epsilon},\, 0<\epsilon<1)$ to present several new types of IST integrable fractional nonlinear evolution equations (fNLEEs) such as the fractional NLS (fNLS), fractional KdV (fKdV), fractional mKdV (fmKdV), fractional sine-Gordon (fsG), and fractional sinh-Gordon (fshG) equations, and found that they were integrable by the IST with GLM-type integral equations to admit the fractional one-soliton solutions.

When the ultra-short (e.g., 100 fs~\cite{book1,book2}) optical pulse propagation is considered, the higher-order dispersive effects (e.g., third-order dispersion) and nonlinear effects (e.g., self-frequency shift and self-steepening) due to the stimulated Raman scattering can not be neglected~\cite{hnls,hnls2,miha21}. As a result, a generation of the NLS equation called the Hirota equation~\cite{hirota}
\bee\label{mKdV}
\begin{array}{l}
 iq_t +\alpha (q_{xx}+2\nu |q|^2q)+i\beta (q_{xxx}+ 6\nu|q|^2q_x)=0, \quad \alpha,\,\beta\in\mathbb{R},\quad \nu=\pm 1,\quad  (x,t)\in \mathbb{R}^2
\end{array}
\ene
was presented, which is also an important physical model, and IST integrable, where $\alpha,\, \beta$ stand for the second- and third-order dispersive coefficients, respectively, and $\nu=+1 (-1)$ corresponds to the self-focusing (defocusing) interaction.
The rogue wave solutions of the focusing Hirota equation were found using the Darboux transform~\cite{hrw1, hrw2, yan13, yanchaos15}.  At $\beta=0$ and $\alpha=1$, Eq.~(\ref{mKdV}) becomes the NLS equation, while as $\alpha=0,\,\beta=1$, Eq.~(\ref{mKdV}) is the complex modified KdV (cmKdV) equation
\bee \label{cmkdv}
\begin{array}{l}
 q_t +q_{xxx}+6\nu |q|^2q_x=0, \quad \nu=\pm 1,\,\, (x,t)\in \mathbb{R}^2,
\end{array}
\ene
In fact, there exist higher-order NLS equation such as the fourth-order Lakshmanan-Porsezian-Daniel (LPD)~\cite{lpd}, and fifth-order NLS equation~\cite{nail2014,Kano89}, and higher-order NLS equations~\cite{Kano89}.

In this paper, we would like to apply the idea (combination of Riesa fractional derivative and IST)~\cite{ab-prl22} to study the integrable fractional Hirota (fHirota) and fractional cmKdV (fcmKdV) equations, and solve them by the IST with the matrix Riemann-Hilbert problems (not the GLM integral equations) such that we find their fractional $N$-soliton solutions. The rest of this paper is arranged as follows. In Sec. II, we give the integrable fHirota and fcmKdV equations, and their explicit forms in terms of the completeness relation of squared eigenfunctions. Similarly, the general higher-order fractional NLS equations are also presented, such as the fractional  LPD equation and fractional fifth-order NLS equations. The anomalous dispersive relations are presented. Moreover, the IST with the matrix Riemann-Hilbert problem for the simple pole case is used to study their fractional solutions. The trace formulae are also studied. In Sec. 3, we give the formula of fractional $N$-soliton solutions for the reflectionless case. Some representative fractional one-, two- and three-soliton solutions are explored for the fHirota and fcmKdV equations. The wave, group, and phase velocities of these envelope fractional solitons are shown to be related to the power laws
of their amplitudes. These obtained fractional $N$-soliton solutions may be useful to explain the related super-dispersion transport of nonlinear wave phenomena in fractional nonlinear media. Moreover, we also present the corresponding formula for the fractional $N$-soliton solutions of the integrable fractional HONLS equations. Finally, some conclusions and discussions are given in Sec. 4.

\section{Fractional Hirota equation and extensions: IST with RH problem}

\quad {\it The fractional Hirota equation}.---The $2\times 2$-matrix ZS-AKNS scattering problem~\cite{nls1,akns} is given as
\bee
\label{lax-x}
\Phi_x=X\Phi,\quad X(x ,t; k)=-ik\sigma_3+Q,\quad
\sigma_3=\begin{bmatrix}
1&0\\
0&-1
\end{bmatrix},\quad
Q=\begin{bmatrix}
0&q(x, t) \\
r(x, t)&0
\end{bmatrix},
\ene
where $\Phi=\Phi(x,t;k)$ is a 2$\times$2 matrix-valued eigenfunction, $k\in \mathbb{C}$ is a spectral parameter, and $r(x,t),\, q(x, t)$ stand for the potentials. Staring from the spectral problem (\ref{lax-x}), one can find the integer-order integrable AKNS hierarchy~\cite{nls1,akns}. Similarly, we here present the coupled fractional Hirota equations
\bee \label{hirota}
 {\bf q}_t+\sigma_3{\cal N}(\widehat{\bf L}){\bf q}=0,\qquad {\bf q}=(r,\, q)^T,
\ene
where $r=r(x,t),\, q=(x,t)$,\, ${\cal N}(\widehat{\bf L})=-4i(\alpha+2\beta \widehat{\bf L})\widehat{\bf L}^2|4\widehat{\bf L}^2|^{\epsilon}  $ with $\epsilon\in (0, 1),$\,
$\alpha,\,\beta\in\mathbb{R}$ and
\bee \sigma_3=\begin{bmatrix}
1&0\\
0&-1
\end{bmatrix},\quad
 \widehat{\bf L}=\frac{1}{2i}\begin{bmatrix} \partial-2r\partial_{-}^{-1}q &  2r\partial_{-}^{-1}r \v\\
         -2q\partial_{-}^{-1}q &  -\partial+2q\partial_{-}^{-1}r \end{bmatrix}
\ene
with $\partial=\partial/\partial x,\, \partial_{-}^{-1}=\int_{-\infty}^xdy$. In particular, as $r=-\nu q^*,\, \nu=\pm 1 $, where the star denotes the complex conjugate, we have the fHirota equation
\bee\label{h2}
 i\begin{bmatrix} -\nu q^* \\ q \end{bmatrix}_t-\sigma_3|4\widehat{\bf L}^2|^{\epsilon}\begin{bmatrix}
-\alpha(\nu q^*_{xx}+2|q|^2q^*)+i\beta (\nu q^*_{xxx}+ 6|q|^2q^*_x) \v\\
\alpha(q_{xx}+2\nu |q|^2q)+i\beta (q_{xxx}+ 6\nu|q|^2q_x)
\end{bmatrix}=0.
\ene
In particular, when $\beta=0$, one has the fractional NLS (fNLS) equation~\cite{ab-prl22}
\bee\label{fnls}
 i\begin{bmatrix} \mp q^* \\ q \end{bmatrix}_t-\sigma_3|4\widehat{\bf L}^2|^{\epsilon}\begin{bmatrix}
-\alpha(\nu q^*_{xx}+2|q|^2q^*) \v\\
\alpha(q_{xx}+ 2\nu|q|^2q)
\end{bmatrix}=0.
\ene
As $\alpha=0,\, \beta=1$, one has the fractional complex mKdV (fcmKdV) equation
\bee\label{m2}
 \begin{bmatrix} -\nu q^* \\ q \end{bmatrix}_t-\sigma_3|4\widehat{\bf L}^2|^{\epsilon}\begin{bmatrix}
-\nu q^*_{xxx}- 6|q|^2q^*_x \v\\
q_{xxx}+6\nu|q|^2q_x
\end{bmatrix}=0.
\ene

{\it Anomalous dispersive relation}.---The formal solution $q(x,t)\sim e^{i[kx-w(k)t]}$ is employed into the linearization of Eq.~(\ref{hirota}) yields the dispersive relation of the linear fHirota equation
\bee
     w(k)=i{\cal N}(-k/2).
\ene
We further consider the linearization of the fHirota Eq.~(\ref{h2})
\bee \no
 iq_t+|\!\!-\!\partial^2|^{\epsilon}(\alpha q_{xx}+i\beta q_{xxx})=0,
\ene
where $|\!\!-\!\partial^2|^{\epsilon}$ denotes the Riesz fractional derivative, such that  the anomalous dispersive relation is
\bee
 w(k)=(\alpha -\beta k)k^2|k^2|^{\epsilon},
\ene
in which we have
 \bee \label{N}
  {\cal N}(k)=-iw(-2k)=-4i(\alpha+2\beta k)k^2|4k^2|^{\epsilon}.
 \ene

{\it Fractional higher-order NLS equations and anomalous dispersive relations}.---In fact, one can also extend the fHirota equation to other fractional higher-order NLS (fHONLS) equations in the form
\bee \label{fnlss}
{\bf q}_t+\sigma_3{\cal N}_h(\widehat{\bf L}){\bf q}=0,\quad {\cal N}_h(\widehat{\bf L})=\left(\sum_{j=2}^N\alpha_j\delta_j\widehat{\bf L}^j\right)|4\widehat{\bf L}^2|^{\epsilon},\quad {\bf q}=(r,\, q)^T,
 \ene
 where $\alpha_j\in\mathbb{R}$ and
\bee \no
 \delta_j=\left\{\begin{array}{ll} i(2i)^j,  & j=2n,\,\,\,\,\, n\in\mathbb{N}, \v\\ (2i)^j, & j=2n+1,\,\,\,\, n\in\mathbb{N}.  \end{array}\right.
\ene
Let $r=-\nu q^*,\, \nu=\pm 1 $, then it follows from Eq.~(\ref{fnlss}) that we have the fHONLS equation
\bee\label{hnls}
 i\begin{bmatrix} -\nu q^* \\ q \end{bmatrix}_t+i\sigma_3|4\widehat{\bf L}^2|^{\epsilon}
 \left(\sum_{j=2}^N\alpha_j\delta_j\widehat{\bf L}^j\right)\begin{bmatrix}-\nu q^* \\ q \end{bmatrix}=0.
\ene
In particular, as $N=5$, we have the fractional fifth-order NLS equation
\bee\label{h5}
 i\begin{bmatrix} \mp q^* \\ q \end{bmatrix}_t-\sigma_3|4\widehat{\bf L}^2|^{\epsilon}\begin{bmatrix}
\alpha_2(\mp q^*_{xx}-2|q|^2q^*)-i\alpha_3 (\mp q^*_{xxx}- 6|q|^2q^*_x)\mp\alpha_4S_4^*[q]\pm i\alpha_5S_5^*[q] \v\\
\alpha_2(q_{xx}\pm 2|q|^2q)+i\alpha_3 (q_{xxx}\pm 6|q|^2q_x)+\alpha_4S_4[q]+i\alpha_5S_5[q]
\end{bmatrix}=0
\ene
with \bee\no
\begin{array}{l}
S_4[q]=q_{xxxx}\pm 8|q|^{2}q_{xx}+6|q|^{4}q \pm 4|q_{x}|^{2}q\pm 6q^*q_{x}^{2}\pm 2q^{2}q_{xx}^*,\v\\
S_5[q]=q_{5x}\pm 10|q|^{2}q_{xxx}\pm 10(q|q _{x}|^{2})_{x}\pm 10q^*(q_{x}^2)_{x}+30|q|^{4}q_{x}.
\end{array}
\ene
which reduce to the fractional Lakshmanan-Porsezian-Daniel (fLPD) equation as $S_5=0$.

We further use fhe formal solution $q(x,t)\sim e^{i[kx-w_h(k)t]}$ to study the linearization of the fHONLS equation (\ref{hnls})
\bee \no
 iq_t+|\!\!-\!\partial^2|^{\epsilon}\left(\sum_{j=2}^N\alpha_j i^{\varsigma_j}q_{jx}\right)=0,\quad
\varsigma=0,\,\, j=2n;\,\, \varsigma=1,\,\, j=2n+1,\,\, n\in \mathbb{N},
\ene
where $q_{jx}=\partial^jq/\partial x^j$, such that  the dispersive relation is
\bee
 w_h(k)=-\sum_{j=2}^N\alpha_ji^{\varsigma_j+j}k^j|k^2|^{\epsilon},
\ene
in which we have
 \bee \label{Ng}
  {\cal N}_h(k)=-iw_h(-2k)=\sum_{j=2}^N\alpha_ji^{\varsigma_j+j+1}(-2k)^j|4k^2|^{\epsilon}.
 \ene

In the following we mainly consider the fHirota and fcmKdV equations (\ref{h2}). In fact, one can also consider the fractional higher-order NLS equations (\ref{hnls}).

{\it The direct scattering with sufficient decay and smoothness of ${\bf q}$}.---For the given ZS-AKNS spectral problem (\ref{lax-x}), following the idea~\cite{ab-prl22}, we now consider the time evolution of the matrix eigenfunction $\Phi(x,t; k)$ in the form
 \bee\label{lax-t}
\begin{array}{ll}
\Phi_t=T\Phi, & \quad
T(x,t; k)=\begin{bmatrix}
  T_1(x,t; k) & T_2(x,t; k) \v\\  T_3(x,t; k) & -T_1(x,t; k)
\end{bmatrix},
\end{array}
\ene
where $T_{1,2,3}(x,t; k)$ can not be explicitly presented in general for the fractional Hirota equation. But we here require  the
conditions $T_{2,3}(x,t; k)\to 0$ and $T_1(x,t; k)\to \frac12 {\cal N}(k)$ with ${\cal N}(k)$ given by Eq.~(\ref{N}) as $x\to \pm \infty$\, (i.e., $q(x,t)\to 0)$. Therefore, for the zero-boundary condition $q(x,t)\in L^1(\mathbb{R}^{\pm})$ and $r=-q^*$, we consider the asymptotic problem $(x\to \pm \infty)$ of the spectral problem (\ref{lax-x}) and time part (\ref{lax-t})
\bee\no
\begin{array}{ll}
\Phi_x=X_{\pm}(k)\Phi, & \quad X_{\pm}(k)=\displaystyle\lim_{x\to\pm \infty}X(x,t; k)=-ik\sigma_3, \v\\
\Phi_t=T_{\pm}(k)\Phi, & \quad T_{\pm}(k)=\displaystyle\lim_{x\to\pm \infty}T(x,t; k)=\frac12 {\cal N}(k)\sigma_3,
\end{array}
\ene
which can deduce the fundamental solutions
$\Phi_{bg}(x, t; k)=e^{[-ikx+\frac12 {\cal N}(k)t]\sigma_3}$.
One may introduce the Jost solutions $\varPhi_{\pm}(x, t; k)$ satisfying the following boundary conditions
\begin{align}\label{Jost-asy}
\Phi_{\pm}(x, t; k)\sim e^{[-ikx+\frac12 {\cal N}(k)t]\sigma_3},\quad x\to\pm\infty.
\end{align}
such that the modified Jost solutions $\phi_{\pm}(x,t;k)$ are defined as
\bee \label{bianjie0}
\phi_{\pm}(x,t;k)=\Phi_{\pm}(x,t;k)e^{(ikx-\frac12{\cal N}(k)t)\sigma_3}\to \mathbb{I},\quad x\to\pm\infty
\ene
with
\begin{align}
\label{Jost-int}
\begin{aligned}
\phi_{\pm}(x,t;k)=\mathbb{I}+\int_{\pm\infty}^xdx'e^{-ik(x-x')\hat{\sigma}_3}Q(x',t)\phi_{\pm}(x',t;k),\quad e^{\hat{\sigma}_3}A=e^{\sigma_3}Ae^{-\sigma_3}.
\end{aligned}
\end{align}

Let $D_+=\{k|\mathrm{Im} ~k>0\},\,\, D_-=\{k|\mathrm{Im} ~k<0\}$, $\Phi_{\pm}(x, t; k)=(\Phi_{\pm 1},\, \Phi_{\pm 2})$ and $\phi_{\pm}(x, t; k)=(\phi_{\pm 1},\, \phi_{\pm 2})$. Then one has the following conclusion (similarly to the NLS equation~\cite{nls1,nls5} and Hirota equation~\cite{zhang20}): For the given $q(x,t)\in L^1\!\left(\mathbb{R^{\pm}}\right)$, the vector-valued functions $\Phi_{\pm 2}$ and $\phi_{\pm 2}$ given by Eqs.~(\ref{bianjie0}) and (\ref{Jost-int}) both have unique solutions in $\mathbb{R}$.  Moreover,
$\phi_{-1,+2},\, \Phi_{-1, +2}$ ($\phi_{-2, +1}$, $\Phi_{-2, +1}$) can be extended analytically to $D_{+}$ ($D_{-}$), and continuously to $D_{+}\cup \mathbb{R}$ ($D_{-}\cup \mathbb{R}$).
Because $\Phi_{\pm}(x, t; k)$ are both fundamental solutions of the spectral problem, thus based on the above properties,
there exists a constant scattering matrix $S(k)=\left(s_{ij}(k)\right)_{2\times 2}$ between them obeying the relation
\bee\label{sr}
 \Phi_+(x,t;k)=\Phi_-(x,t;k)S(k),\quad  k\in\mathbb{R},
 \ene
which can generate $s_{11}s_{22}-s_{12}s_{21}=1$ and
 \begin{align}\label{S-lie}
\begin{aligned}
s_{11}(k)=\left|\Phi_{+1}(x, t; k), \Phi_{-2}(x, t; k)\right|,\quad s_{22}(k)=|\Phi_{-1}(x, t; z), \Phi_{+2}(x, t; z)|,\\[0.05in]
s_{12}(k)=|\Phi_{+2}(x, t; k), \Phi_{-2}(x, t; k)|,\quad s_{21}(k)=|\Phi_{-1}(x, t; k), \Phi_{+1}(x, t; k)|.
\end{aligned}
\end{align}
According to the properties of $\Phi_{\pm j}(x, t; k),\, j=1,2$, it can be seen that the scattering coefficient $s_{11}(k)$
($s_{22}(k)$) in $k\in\mathbb{R}$ can be extended analytically to $D_{-}$ ($D_{+}$), and continuously to $D_{-}\cup \mathbb{R}$
($D_{+}\cup \mathbb{R}$), whereas another two scattering coefficients $s_{12}(k)$ and $s_{21}(k)$ can not be analytically continued away from $\mathbb{R}$. To study the matrix RH problem of the inverse scattering, the potential is required not to admit spectral singularities~\cite{Zhou1989}, i.e. $s_{11}(k)s_{22}(k)\ne 0$ as $k\in\mathbb{R}$ such that the reflection coefficients are introduced as $\rho(k)=s_{21}(k)s^{-1}_{11}(k),\, \hat\rho(z)=s_{12}(k)s^{-1}_{22}(k)$ as $k\in \mathbb{R}.$

{\it Completeness of squared eigenfunctions and explicit fHirota equation and higher-order extensions}.---Since $\widehat{\bf L}$ is a adjoint of the matrix operator ${\bf L}$
\bee {\bf L}=\frac{1}{2i}\begin{bmatrix}
      -\partial-2q\partial_{+}^{-1}r &  -2q\partial_{+}^{-1}q \v\\
         2r\partial_{+}^{-1}r &  \partial+2r\partial_{+}^{-1}q \end{bmatrix}, \qquad \partial_{+}^{-1}=\int_x^{\infty} dy.
\ene
It follows from the ZS-AKNS spectral problem that the eigenfunctions ${\Psi}_{+j},\, \widehat{\Psi}_{-j}$ of ${\bf L}$ and $\widehat{\bf L}$ and eigenvalue $k$ satisfy~\cite{akns,kaup76}
\bee
\begin{array}{l}
 {\bf L}{\Psi}_{+j}=k{\Psi}_{+j},\quad \widehat{\bf L}\widehat{\Psi}_{-j}=k\widehat{\Psi}_{-j}, \quad j=1,2,\v\\
 \Psi_{+1}=(\Phi^2_{+11}(x, t; k),\,  \Phi_{+12}^2(x, t; k))^T, \qquad
 {\Psi}_{+2}=(\Phi_{+21}^2(x, t; k), \,  \Phi_{+22}^2(x, t; k))^T, \v \\
 \widehat{\Psi}_{-1}=(\Phi_{-12}^2(x, t; k),\,  -\Phi_{-11}^2(x, t; k))^T, \quad
 \widehat{\Psi}_{-2}=(\Phi_{-22}^2(x, t; k), \, -\Phi_{-21}^2(x, t; k))^T,
 \end{array}
 \ene
and they are complete.

Similarly,
\bee
{\cal N}(\widehat{\bf L})\widehat{\Psi}_{-j}={\cal N}(k)\widehat{\Psi}_{-j},\quad j=1,2.
\ene
Since the eigenfunctions $\widehat{\Psi}_{-j}$ are complete, therefore, one can use
${\cal N}(\widehat{\bf L})$ to act on a sufficiently smooth and decaying vector function ${\bf g}(x)=(g_1(x),\, g_2(x))^{\mathrm{T}}$ to get
\bee\label{jifen1}
{\cal N}(\widehat{\bf L}){\bf g}(x)=\frac{1}{\pi}\sum\limits_{\ell=1}^2\int_{\Gamma_{\infty,\ell}}dk
{\cal N}(k)f_{\ell}(k)\int_{\mathbb{R}}G_{\ell}(x,y,k){\bf g}(y)dy,
\ene
where
\begin{align} \no
\begin{array}{l}
G_1(x,y,k)=\widehat{\Psi}_{-1}(x,k)\Psi_{+2}^{\mathrm T}(y,k),\quad f_1(k)=-s_{22}^{-2}(k),\quad k\in D_+, \v\\
G_2(x,y,k)=\widehat{\Psi}_{-2}(x,k)\Psi_{+1}^{\mathrm T}(y,k),\quad f_2(k)=s_{11}^{-2}(k),\quad k\in D_-,
\end{array}
\end{align}
and $\Gamma_{\infty, \ell}=\lim\limits_{\mathbb{R}\to\infty}\Gamma_{\mathbb{R}, \ell},\, \ell=1,2$ with
$\Gamma_{\mathbb{R}, 1}$ ($\Gamma_{\mathbb{R}, 2}$) the semicircular contour in the upper (lower) half plane evaluated $-\mathbb{R}$ to $\mathbb{R}$.

Then, the operation of ${\cal N}(\widehat{\bf L})=-4i(\alpha+2\beta\widehat{\bf L})\widehat{\bf L}^2|4\widehat{\bf L}^2|^\epsilon$  on the ${\bf q}=(-q^*, q)^T$ yields
\bee\label{jifen2}
-4i(\alpha+2\beta\widehat{\bf L})\widehat{\bf L}^2|4\widehat{\bf L}^2|^\epsilon\bm q=\frac{1}{\pi}\sum\limits_{\ell=1}^2\int_{\Gamma_{\infty,\ell}}dk|4k^2|^\epsilon f_{\ell}(k)\int_{\mathbb{R}}dyG_{\ell}(x,y,k)\begin{bmatrix}
i\alpha (q_{yy}^*+2|q|^2q^*)-\beta (q_{yyy}^*+6 |q|^2q_y^*)\\[0.05in]
i\alpha (q_{yy}+2|q|^2q)-\beta (q_{yyy}+6 |q|^2q_y)
\end{bmatrix}.
\ene

Therefore it follows from Eq.~(\ref{jifen2}) and Eq.~(\ref{h2}) with $\nu=1$ that we have the explicit representation of the
fHirota equation
\bee\no
 i\begin{bmatrix} - q^* \\ q \end{bmatrix}_t-\frac{\sigma_3}{\pi}
 \sum\limits_{{\ell}=1}^2\int_{\Gamma_{\infty}^{\ell}}|4k^2|^\epsilon f_{\ell}(k)dk\int_{\mathbb R}G_{\ell}(x,y,k)\begin{bmatrix}
-\alpha(q^*_{yy}+2|q|^2q^*)+i\beta (q^*_{yyy}+ 6|q|^2q^*_y) \v\\
\alpha(q_{yy}+ 2|q|^2q)+i\beta (q_{yyy}+ 6|q|^2q_y)
\end{bmatrix}dy=0,
\ene
that is the fHirota equation is given by
\bee\label{fhirota-exp}
 iq_t+\frac{1}{\pi}\sum\limits_{{\ell}=1}^2\int_{\Gamma_{\infty}^{\ell}}dk|4k^2|^\epsilon f_{\ell}(k)
  \int_{\mathbb{R}}{\cal G}_{\ell}(x,y,k)dy=0,
  \ene
where
 \bee\no
 \begin{array}{rl}
{\cal G}_j(x,y,k)=& -\Phi_{-j1}^2(x,k)\Big\{\Phi_{+(3-j)1}^2(y,k)[-\alpha(q^*_{yy}+2|q|^2q^*)+i\beta (q^*_{yyy}+ 6|q|^2q^*_y)]\v\\
  & \,\, +\Phi_{+(3-j)2}^2(y,k)[\alpha(q_{yy}+ 2|q|^2q)+i\beta (q_{yyy}+ 6|q|^2q_y)]\Big\},\quad j=1,2.
\end{array}
\ene
As $\alpha=0,\, \beta=1$, Eq.~(\ref{fhirota-exp}) reduces to the explicit  fcmKdV equation. As $\alpha=1,\, \beta=0$, Eq.~(\ref{fhirota-exp}) reduces to the explicit fNLS equation. Moreover, at $\epsilon=0$, one can recover the focusing Hirota equation (\ref{mKdV}) with $\nu=1$ from Eq.~(\ref{fhirota-exp}).

Similarly, these results can also be extended to the higher-order case (\ref{hnls}). Let
\bee \no
\begin{bmatrix}
\widehat{q}_1(x,t; \alpha_j, \epsilon) \v\\ \widehat{q}_2(x,t; \alpha_j, \epsilon)
\end{bmatrix}=\left(\sum_{j=2}^N\alpha_j\delta_j\widehat{\bf L}^j\right)\begin{bmatrix} -q^* \v\\ q \end{bmatrix}.
 \ene
Then, according to the above-mentioned idea, we have the fractional HONLS equation (\ref{fnlss}) in the explicit form
\bee
 iq_t+\frac{1}{\pi}\sum\limits_{{\ell}=1}^2\int_{\Gamma_{\infty}^{\ell}}dk|4k^2|^\epsilon f_{\ell}(k)
  \int_{\mathbb{R}}\widehat{\cal G}_{\ell}(x,y,k)dy=0,
  \ene
where
 \bee\no
 \begin{array}{rl}
\widehat{\cal G}_j(x,y,k)=& -\Phi_{-j1}^2(x,k)\Big\{\Phi_{+(3-j)1}^2(y,k)\widehat{q}_1(y,t; \alpha_j, \epsilon)+\Phi_{+(3-j)2}^2(y,k)\widehat{q}_2(y,t; \alpha_j, \epsilon)\Big\},\quad j=1,2.
\end{array}
\ene

{\it Inverse scattering: symmetry, discrete spectrum, and residue conditions}.---We consider the symmetry properties of scattering matrix as follows:
\bee\label{ss}
S(k)=\sigma S^*(k^*)\sigma^{-1},\qquad \sigma=
\begin{bmatrix}
0&1 \\
-1&0
\end{bmatrix}.
\ene

The continuous spectrum set of $X_{\pm}$ is  $k\in \mathbb{R}$, which is the jump contour for the following considered Riemann-Hilbert (RH) problem. The discrete spectrum of the scattering problem is the set of $k\in \mathbb{C}\backslash \mathbb{R}$ such that they admit eigenfunctions in $L^2(\mathbb{R})$. For the fHirota equation (\ref{h2}) or fHONLS equation (\ref{hnls}), they should satisfy $s_{jj}(k_j)=0,\, j=1,2$ for
$k_1\in D_-,\, k_2\in D_+$. We assume that $s_{22}(k)$ has $N$ simple zeros in $D_+$ denoted by $k_n$, $n=1, 2, \cdots, N$, that is, $s_{22}(k_n)=0$ and $s_{22}'(k_n)\not=0\, (n=1,2,...,N)$. It follows from Eq.~(\ref{ss}) that if $s_{22}(k_n)=0$, then $s_{11}(k_n^*)=0$. The discrete spectrum set is
$K=\{\{k_n,\, k_n^*\} | s_{22}(k_n)=0,\, k_n\in D_+\}$.
Since $s_{22}(k_0)=0$,\, $s_{22}'(k_0)\not=0$,\, $s_{11}(k_0^*)=0$ and $s_{11}'(k_0^*)\not=0$ are required for $k_0\in D_+,\, k_0^*\in D_-$, then it follows from Eq.~(\ref{S-lie}) that there exist two norming constants $b_{\pm}(k_0)$ satisfying
\begin{align} \no
\Phi_{+2}(x, t; k_0)=b_+(k_0)\,\Phi_{-1}(x, t; k_0),\quad
\Phi_{+1}(x, t; k_0^*)=b_-(k_0^*)\,\Phi_{-2}(x, t; k_0^*),
\end{align}
which can generate the residue conditions
\bee\label{s1}
\begin{array}{l}
\d\mathop\mathrm{Res}\limits_{k=k_0}\left[\frac{\Phi_{+2}(x, t; k)}{s_{22}(z)}\right]=\frac{\Phi_{+2}(x, t; k_0)}{s_{22}'(k_0)}=\frac{b_+(k_0)}{s_{22}'(k_0)}\,\Phi_{-1}(x, t; k_0)=A_+[k_0]\,\Phi_{-1}(x, t; k_0), \v\\
\d\mathop\mathrm{Res}\limits_{k=k_0^*}\left[\frac{\Phi_{+1}(x, t; k)}{s_{11}(k)}\right]=\frac{\Phi_{+1}(x, t; k_0^*)}{s_{11}'(k_0^*)}=\frac{b_-(k_0^*)}{s_{11}'(k_0^*)}\,\Phi_{-2}(x, t; k_0^*)=A_-[k_0^*]\,\Phi_{-2}(x, t; k_0^*).
\end{array}
\ene
Moreover, one has the symmetry $A_+[k_0]=-A_-^*[k_0^*],\quad k_0\in D_+,\,\, k_0^*\in D_-.$

{\it Riemann-Hilbert problem}.---Let the sectionally meromorphic matrix $M(x, t; k)$ be
\begin{align}\no
M(x, t; k)=\left\{
\begin{aligned}
\left(\phi_{-1}(x, t; k),\, \frac{\phi_{+2}(x, t; k)}{s_{22}(k)}\right),\quad k\in D_+, \\[0.05in]
\left(\frac{\phi_{+1}(x, t; k)}{s_{11}(k)},\, \phi_{-2}(x, t; k)\right), \quad k\in D_-.
\end{aligned}\right.
\end{align}
where $M^{\pm}(x, t; k)=\displaystyle\lim_{\tilde{k}\to k,\, \tilde{k}\in D_{\pm} }M(x, t; \tilde{k}), \quad k\in\mathbb{R}.$
Then according to the properties of $\phi_{\pm j}(x, t; k),\, j=1,2$ and
\begin{align}\no
\left(\Phi_{-1}(x, t; k), \frac{\Phi_{+2}(x, t; k)}{s_{22}(k)}\right)=\left(\frac{\Phi_{+1}(x, t; k)}{s_{11}(k)}, \Phi_{-2}(x, t; k)\right)(\mathbb{I}-V_0(k)),\quad V_0(k)=\left(\begin{array}{cc} 0 & -\hat\rho(k)\\[0.05in] \rho(k)& \rho(k)\hat\rho(k) \end{array}\right),
\end{align}
one has the following Riemann-Hilbert (RH) problem satisfied by the matrix function $M(x, t; k)$:
 {\it \begin{itemize}
 \item {} Analyticity: $M(x, t; k)$ is analytic in $D_{\pm}\setminus K$, and has the simple poles in $K$;

 \item {} Jump relation: $M^-(x, t; k)=M^+(x, t; k)\left(\mathbb{I}-V(x, t; k)\right)$ with
$V(x, t; k)=e^{(ikx-\frac12{\cal N}(k)t)\hat{\sigma}_3}V_0^{-1}(k)$,\, $k\in R$;
\item {} Asymptoticity: $M^{\pm}(x, t; k)\to \mathbb{I}$ for $k\to\infty$.
\end{itemize} }

Considering the asymptotic behavior and the simple-pole contributions (cf. Eq.~(\ref{s1})] of $M(x, t; k)$
\bee \label{msp}
\begin{array}{rl}
M_{sp}(x, t; k)=&\d\mathbb{I}+\sum_{n=1}^{N}\left[\frac{\mathop\mathrm{Res}\limits_{k=k_n}M^+(x, t; k)}{k-k_n}+\frac{\mathop\mathrm{Res}\limits_{k=k_n^*}M^-(x, t; k)}{k-k_n^*}\right] \v\\
=&\d \mathbb{I}+\left[\dfrac{A_-[k_n^*]\,\mathrm{e}^{-2i\theta_{\epsilon}(x, t; k_n^*)}}{k-k_n^*}\,\phi_{-2}(x, t; k_n^*),\dfrac{A_+[k_n]\,\mathrm{e}^{2i\theta_{\epsilon}(x, t; k_n)}}{k-k_n}\,\phi_{-1}(x, t; k_n)\right],
\end{array}
\ene
where $\theta_{\epsilon}(x,t;k)=-kx+\frac{1}{2i}{\cal N}(k)t=-kx-2(\alpha+2\beta k)k^2|4k^2|^{\epsilon}t$ and then subtracting out $M_{sp}(x, t; k)$ from both sides of the above-given jump relation can generate
\begin{align}\label{jumpbianxing}
\begin{aligned}
&M^-(x, t; k)-M_{sp}(x, t; k)=M^+(x, t; k)-M_{sp}(x, t; k)-M^+(x, t; k)\,V(x, t; k),
\end{aligned}
\end{align}
where $M^{\pm}(x, t; k)-M_{sp}(x, t; k)$ are analytic in $D_{\pm}$. The Cauchy projectors $P^{\pm}\left[f\right](k)=\frac{1}{2\pi i}\int_{R}\frac{f(\zeta)}{\zeta-(k\pm i0)}\,d\zeta$
(where the signs $k\pm i0$ denotes the limit chosen from the left/right of $k$) and Plemelj's formulae are used
to solve Eq.~(\ref{jumpbianxing}) to find the solution of the RH problem
\begin{align}\label{RHP-jie}
\begin{aligned}
M(x, t; k)=M_{sp}(x, t; k)+\frac{1}{2\pi i}\int_{R}\frac{M^+(x, t; \zeta)\,V(x, t; \zeta)}{\zeta-k}\,\mathrm{d}\zeta,\quad k\in\mathbb{C}\backslash R.
\end{aligned}
\end{align}

As $k= k_s\, (k_s^*),\, s=1, 2, \cdots, N$, it follows from the first (second) column of $M(x,t;k)$ given by Eq.~(\ref{RHP-jie}) with Eq.~(\ref{msp}) that
\bee\label{sys}
\begin{array}{l}
\d\phi_{-1}(x, t; k_s)=
\begin{bmatrix}
1\\[0.05in]0
\end{bmatrix}
+\sum_{n=1}^{N}\frac{A_-[k_n^*]\,\mathrm{e}^{-2i\theta_{\epsilon}(x, t; k_n^*)}}{k_s-k_n^*}\,
\phi_{-2}(x, t; k_n^*)+\frac{1}{2\pi i}\int_{R}\frac{\left(M^+V\right)_1(x, t; \zeta)}{\zeta-k_s}\,d\zeta,\quad s=1, 2, \cdots, N, \v\\
\d\phi_{-2}(x, t; k_s^*)=
\begin{bmatrix}
0\\[0.05in]1
\end{bmatrix}
+\sum_{n=1}^{N}\frac{A_+[k_n]\,\mathrm{e}^{2i\theta_{\epsilon}(x, t; k_n)}}{k_s^*-k_n}\,
\phi_{-1}(x, t; k_n)+\frac{1}{2\pi i}\int_{R}\frac{\left(M^+V\right)_2(x, t; \zeta)}{\zeta-k_s^*}\,d\zeta,\quad s=1, 2, \cdots, N.
\end{array}
\ene


According to the condition $q(x, t)=\lim_{k\to\infty}(kM)_{12}$ and the solutions $\phi_{-11}(x, t; k_n),\, n=1,2,...,N)$ given by system (\ref{sys}), we have the factional simple-pole solutions of the fractional Hirota equation as
\bee\label{solution}
q(x, t)=2i\sum_{n=1}^{N}A_+[k_n]\,\mathrm{e}^{2i\theta_{\epsilon}(x, t; k_n)}\,\phi_{-11}(x, t; k_n)+
\frac{1}{2\pi i}\int_{R}\left(M^+V\right)_{12}(x, t; \zeta)\,d\zeta.
\ene

 Since the scattering coefficients $s_{22}(k)$ and $s_{11}(k)$ are analytic in $D_+$ and $D_-$, respectively, and the discrete spectral points $k_n$'s and $k_n^*$'s are the simple zeros of $s_{22}(k)$ and $s_{11}(k)$, respectively, then one can also find the trace formulae for the fHirota equation
\begin{align}\no
s_{22}(k)=e^{s(k)}s_0(k) \,\, {\rm for}\,\, k\in D_+, \quad s_{11}(k)=e^{-s(k)}/s_0(k)\,\, {\rm for}\,\, k\in D_-,
\end{align}
where $s(k)=\frac{i}{2\pi}\int_{\mathbb{R}}\frac{\log\left[1+\rho(\zeta)\,\rho^*(\zeta^*)\right]}{\zeta-k}\,d\zeta$ and $s_0(k)=\prod_{n=1}^N\frac{(k-k_n)}{(k-k_n^*)}$.\v

\section{Dynamics of fractional $N$-soliton solutions}

\quad {\it Fractional $N$-soliton solutions}.---We consider the case of reflectionless potential (i.e., $V=0$) in system (\ref{sys}) such that we have the system of equations with respect to $\phi_{-11}(x, t; k_n),\, n=1,2,...,N$
\bee\label{sys2}
\left(1-\sum_{n_1=1}^{N}\sum_{n_2=1}^{N}\frac{A_-[k_{n_1}^*]\,\mathrm{e}^{-2i\theta_{\epsilon}(x, t; k_{n_1}^*)}}{k_n-k_{n_1}^*}\,
\frac{A_+[k_{n_2}]\,\mathrm{e}^{2i\theta_{\epsilon}(x, t; k_{n_2})}}{k_{n_1}^*-k_{n_2}}\right)\,
\phi_{-11}(x, t; k_{n_2})=1,\quad n=1, 2, \cdots, N.
\ene
According to the condition $q(x, t)=\lim_{k\to\infty}(kM)_{12}$ and the solutions $\phi_{-11}(x, t; k_n),\, n=1,2,...,N)$ given by system (\ref{sys2}), we have the factional $N$-soliton solutions (reflectionless case) with simple poles of the fHirota equation as
\bee\label{recons}
q(x, t)=2i\sum_{n=1}^{N}A_+[k_n]\,\mathrm{e}^{2i\theta_{\epsilon}(x, t; k_n)}\,\phi_{-11}(x, t; k_n)=-2i\dfrac{\left|\begin{matrix} \mathbb{I}_N+H & {\bf a}^T \vspace{0.05in}\\ {\bf b} &0 \end{matrix}\right|}{|\mathbb{I}_N+H|},
\ene
where $\mathbb{I}_N$ is an $N\times N$ unit matrix, $\theta_{\epsilon}(x, t; k)=-kx-2(\alpha+2\beta k)k^2|4k^2|^{\epsilon}t$,
\begin{align}\no
\begin{aligned}
&{\bf a}=(a_j)_{1\times N}=\left(1,1,..,1\right),\quad {\bf b}=(b_j)_{1\times N}=(b_1, b_2,..,b_N),\quad b_n=A_+[k_n]\mathrm{e}^{2i\theta_{\epsilon}(x, t; k_n)},\quad n=1,2,...,N,\\
&H=(H_{(n,n_2)})_{N\times N},\quad H_{(n,n_2)}=\sum_{n_1=1}^{N}\frac{A_-[k_{n_1}^*]\,\mathrm{e}^{-2i\theta_{\epsilon}(x, t; k_{n_1}^*)}}{k_n-k_{n_1}^*}\,
\frac{A_+[k_{n_2}]\,\mathrm{e}^{2i\theta_{\epsilon}(x, t; k_{n_2})}}{k_{n_2}-k_{n_1}^*}.
\end{aligned}
\end{align}

\begin{figure}[!t]
    \centering
\vspace{0.05in}
  {\scalebox{0.7}[0.7]{\includegraphics{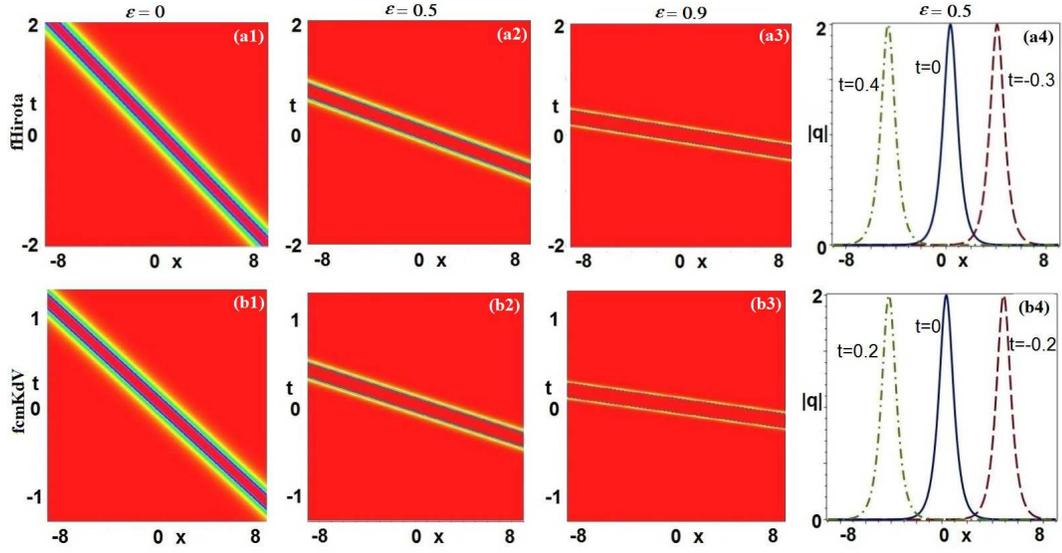}}}\hspace{-0.35in}
\vspace{0.15in}
\caption{Fractional 1-soliton solutions (\ref{solu1}) for $k_1=1+i,\, A_+[k_1]=1$ and $\epsilon=0,\, 0.5,\, 0.9$. (a1)-(a4)
fHirota at $\alpha=1,\, \beta=0.05$; (b1)-(b4) fcmKdV at $\alpha=0,\, \beta=1$.}
   \label{fig1}
\end{figure}

In what follows we will discuss some fractional one- two-, and three-soliton solutions of the fHirota, fNLS and fcmKdV equations in terms of solution (\ref{recons}) with $N=1, 2, 3$.

\v {\it Fractional one-soliton solutions and velocities}.---For $N=1$, if we choose the spectral parameter $k_1=\xi+i\eta$ and $A_+[k_1]=a$ with $\xi,\, \eta\in\mathbb{R}$ with $a\eta\not=0$, then the expression of fractional 1-soliton solution of the fHirota equation can be written as:
\bee\label{solu1}
 q(x,t)\!=\!2\eta{\rm sech}\Big[2\eta x\!+\!8\eta\big(\alpha \xi\!+\!\beta(3\xi^2\!\!-\!\!\eta^2)\big)|2k_1|^{2\epsilon}t\!-\!\ln|2\eta\!/\!a|\!\Big]e^{-2i\xi x+4i[\alpha(\eta^2-\xi^2)+2\beta \xi(3\eta^2-\xi^2)]|2k_1|^{2\epsilon}t+i(\pi/2+\vartheta)},
\ene
where $\vartheta={\rm Arg}(a/2\eta)$.

Particularly, as $\beta=0$, we have the fractional 1-soliton of the fNLS equation. When $\alpha=0$, we have the fractional 1-soliton of the fcmKdV equation. It follows from the solution (\ref{solu1}) that the wave velocity $v_{wv}(\xi, \eta, \epsilon; \alpha, \beta)$, group velocity $v_{gv}(\xi, \eta, \epsilon; \alpha, \beta)$ and phase velocity $v_{pv}(\xi, \eta, \epsilon; \alpha, \beta)$ of the fractional 1-soliton solutions, respectively, are
\bee \label{velo}
\begin{array}{l}
v_{wv}(\xi, \eta, \epsilon; \alpha, \beta)=-4^{1+\epsilon}\big[\alpha \xi\!+\!\beta(3\xi^2\!-\!\eta^2)\big](\xi^2\!+\!\eta^2)^{\epsilon},\v\\
v_{gv}(\xi, \eta, \epsilon; \alpha, \beta)=4^{1+\epsilon}
 \Big\{\epsilon \xi\left[\alpha (\eta^2-\xi^2)\!+\!2\beta\xi(3\eta^2\!-\!\xi^2)\right]
 -(\xi^2\!+\!\eta^2)\left[\alpha \xi\!+\!3\beta(\xi^2\!-\!\eta^2)\!\right]\!\!\Big\}(\xi^2\!+\!\eta^2)^{\epsilon-1}, \v\\
v_{pv}(\xi, \eta, \epsilon; \alpha, \beta)=2^{1+2\epsilon}\!\left[\alpha\xi^{-1}(\eta^2-\xi^2)+2\beta(3\eta^2-\xi^2)\right](\xi^2\!+\!\eta^2)^{\epsilon},
\end{array}
\ene
which depend on the real ($\xi$) and imaginary $(\eta)$ parts of the special parameter $k_1$, fractional parameter $\epsilon$ and equation coefficients $\alpha,\, \beta$.

For the given $\xi=\eta=1$, i.e., $k_1=1+i$, Figs.~\ref{fig1}(a1-a4) display the fractional 1-soliton solution of the fHirota equation as $\alpha=1,\, \beta=0.05$, which are left-going travelling-wave solitons ($v_{wv}=-1.1\times 2^{2+3\epsilon}<0$) at $\epsilon=0,\, 0.5,\, 0.9$. Moreover, the absolute value of left-going travelling-wave velocity becomes larger as $\epsilon$ increases (see Fig.~\ref{fig-line}(a)).  Figs.~\ref{fig1}(b1-b4) display the fractional 1-soliton solution of the fcmKdV equation as $\alpha=0,\, \beta=1$, which are also left-going travelling-wave solitons ($v_{wv}=- 2^{3+3\epsilon}<0$) at $\epsilon=0,\, 0.5,\, 0.9$. In fact, if we take $\alpha \xi\!+\!\beta(3\xi^2\!-\!\eta^2)<0$, then we have the right-going travelling-wave solitons of the fHirota equation with (e.g., $\xi=\eta=1,\, \alpha=-1,\, \beta=-0.05$), fNLS equation (e.g., $\xi=\eta=1,\, \alpha=-1,\, \beta=0$), and fcmKdV equation (e.g., $\xi=\eta=1,\, \alpha=0,\, \beta=-0.05$).  Fig.~\ref{fig-line}(a) displays the wave velocities (see $v_{wv}$ in Eq.~(\ref{velo})) of the fractional 1-solitons solutions of the fHirota, fNLS and fcmKdV equations, where the absolute values of the wave velocities increase as $\epsilon$ grows.
Figs.~\ref{fig-line}(b, c) show the group velocities (see $v_{gv}$ in Eq.~(\ref{velo})) and phase velocities (see $v_{pv}$ in Eq.~(\ref{velo})) of the fractional 1-soliton solutions of the fHirota, fNLS and fcmKdV equations, respectively.

\begin{figure}[!t]
    \centering
\vspace{-0.05in}
  {\scalebox{0.7}[0.7]{\includegraphics{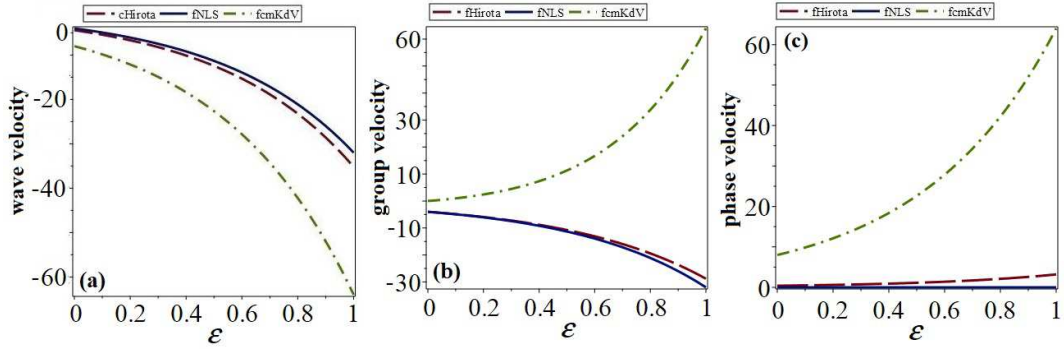}}}\hspace{-0.35in}
\vspace{0.05in}
\caption{(a) wave velocities, (b) group velocities and (c) phase velocities given by Eq.~(\ref{velo}) of fractional 1-soliton solutions of
 fHirota (dashed line, $\alpha=1,\, \beta=0.05$), fNLS (solid line, $\alpha=1,\, \beta=0$) and fcmKdV (dash-dotted line, $\alpha=0,\, \beta=1$) equations at $k_1=1+i,\, A_+[k_1]=1$.}
   \label{fig-line}
\end{figure}

\v {\it Elastic interactions of fractional two-soliton solutions}.---For $N=2$, if we take two spectral parameters $k_1,\, k_2$, and $A_+[k_1]=a,\, A_+[k_2]=b$, then the fractional two-soliton solutions of the fHirota equation are
\bee\label{solu2}
q(x,t)=-2i\frac{A(x,t)}{B(x,t)},
\ene
where
\bee \no
 A(x,t)=\left|\,\begin{matrix} 1-\dfrac{|a|^2e^{2i\theta_{\epsilon}(k_1^*)-2i\theta_{\epsilon}(k_1)}}{(k_1^*-k_1)^2}-\dfrac{a^*be^{2i\theta_{\epsilon}(k_1^*)-2i\theta_{\epsilon}(k_2)}}{(k_1^*-k_2)^2}
 & -\dfrac{ab^*e^{2i\theta_{\epsilon}(k_2^*)-2i\theta_{\epsilon}(k_1)}}{(k_1^*-k_1)(k_2^*-k_1)}
   -\dfrac{|b|^2e^{2i\theta_{\epsilon}(k_2^*)-2i\theta_{\epsilon}(k_2)}}{(k_1^*-k_2)(k_2^*-k_2)}
 & 1 \vspace{0.15in}\\
 -\dfrac{|a|^2e^{2i\theta_{\epsilon}(k_1^*)-2i\theta_{\epsilon}(k_1)}}{(k_1^*-k_1)(k_2^*-k_1)}
 -\dfrac{a^*be^{2i\theta_{\epsilon}(k_1^*)-2i\theta_{\epsilon}(k_2)}}{(k_1^*-k_2)(k_2^*-k_2)}
  &\,\, 1-\dfrac{ab^*e^{2i\theta_{\epsilon}(k_2^*)-2i\theta_{\epsilon}(k_1)}}{(k_2^*-k_1)^2}-\dfrac{|b|^2e^{2i\theta_{\epsilon}(k_2^*)-2i\theta_{\epsilon}(k_2)}}{(k_2^*-k_2)^2}
  & 1 \vspace{0.15in}\\
  -a^*e^{2i\theta_{\epsilon}(k_1^*)} & -b^*e^{2i\theta_{\epsilon}(k_2^*)} & 0
\end{matrix}\,\right|,
\ene
\bee\no
 B(x,t)=\left|\,\begin{matrix} 1-\dfrac{|a|^2e^{2i\theta_{\epsilon}(k_1^*)-2i\theta_{\epsilon}(k_1)}}{(k_1^*-k_1)^2}-\dfrac{a^*be^{2i\theta_{\epsilon}(k_1^*)-2i\theta_{\epsilon}(k_2)}}{(k_1^*-k_2)^2} &
 -\dfrac{ab^*e^{2i\theta_{\epsilon}(k_2^*)-2i\theta_{\epsilon}(k_1)}}{(k_1^*-k_1)(k_2^*-k_1)}-\dfrac{|b|^2e^{2i\theta_{\epsilon}(k_2^*)-2i\theta_{\epsilon}(k_2)}}{(k_1^*-k_2)(k_2^*-k_2)} \vspace{0.15in}\\
 -\dfrac{|a|^2e^{2i\theta_{\epsilon}(k_1^*)-2i\theta_{\epsilon}(k_1)}}{(k_1^*-k_1)(k_2^*-k_1)}-\dfrac{a^*be^{2i\theta_{\epsilon}(k_1^*)-2i\theta_{\epsilon}(k_2)}}{(k_1^*-k_2)(k_2^*-k_2)}
  &\,\,
  1-\dfrac{ab^*e^{2i\theta_{\epsilon}(k_2^*)-2i\theta_{\epsilon}(k_1)}}{(k_2^*-k_1)^2}-\dfrac{|b|^2e^{2i\theta_{\epsilon}(k_2^*)-2i\theta_{\epsilon}(k_2)}}{(k_2^*-k_2)^2}
\end{matrix}\,\right|.
\ene
In particular, as $\beta=0$, we have the fractional two-soliton solutions of the fNLS equation. When $\alpha=0$, we get the fractional two-soliton solutions of the fcmKdV equation.

\begin{figure}[!t]
    \centering
\vspace{-0.05in}
  {\scalebox{0.7}[0.7]{\includegraphics{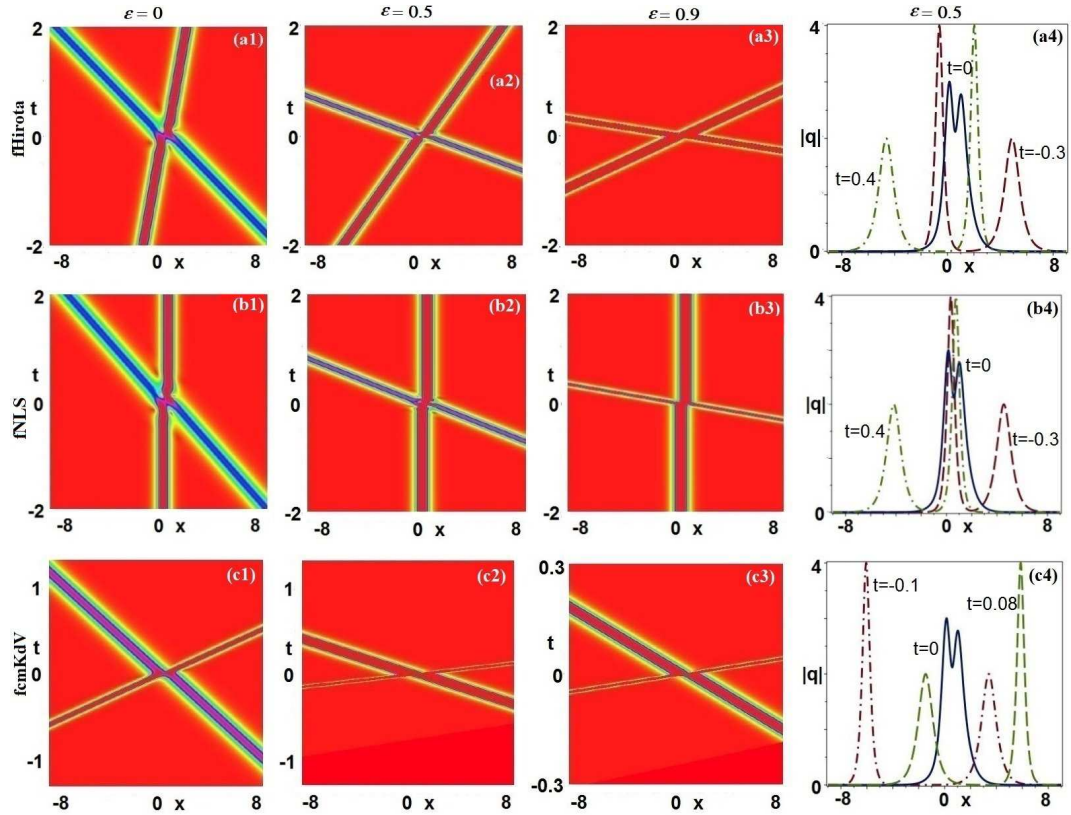}}}\hspace{-0.35in}
\vspace{0.15in}
\caption{Elastic interactions of fractional two-soliton solutions (\ref{solu2}) for $k_1=1+i,\, k_2=2i, \, A_+[k_1]=A_+[k_2]=1$ and $\epsilon=0, 0.5, 0.9$. (a1)-(a4)
fHirota at $\alpha=1,\, \beta=0.05$; (b1)-(b4) fNLS at $\alpha=1,\, \beta=0$; (c1)-(c4) fcmKdV at $\alpha=0,\, \beta=1$.}
   \label{fig2}
\end{figure}

For the two chosen spectral parameters $k_1=1+i,\, k_2=2i$ and $A_+[k_1]=A_+[k_2]=1$, Figs.~\ref{fig2}(a1-a4) display the fractional 2-soliton solution of the fHirota equation as $\alpha=1,\, \beta=0.05$, which imply the elastic interactions of one right-going (left branch) and another left-going (right branch) travelling-wave solitons at $\epsilon=0,\, 0.5,\, 0.9$. In particular, it follows from Figs.~\ref{fig2}(a4) that before the elastic collision, the left-branch is a right-going travelling wave with larger amplitude, and the right branch is a left-going travelling wave with lower amplitude. Figs.~\ref{fig2}(b1-b4) display the fractional two-soliton solutions of the fNLS equation as $\alpha=0,\, \beta=1$, which also imply the elastic interactions of one left-going and another center-going travelling-wave solitons at $\epsilon=0,\, 0.5,\, 0.9$. Figs.~\ref{fig2}(c1-c4) show the fractional two-soliton solutions of the fcmKdV equation as $\alpha=0,\, \beta=1$, which are also elastic interactions of one left-going and another right-going travelling-wave solitons at $\epsilon=0,\, 0.5,\, 0.9$.

\begin{figure}[!t]
    \centering
\vspace{-0.05in}
  {\scalebox{0.7}[0.7]{\includegraphics{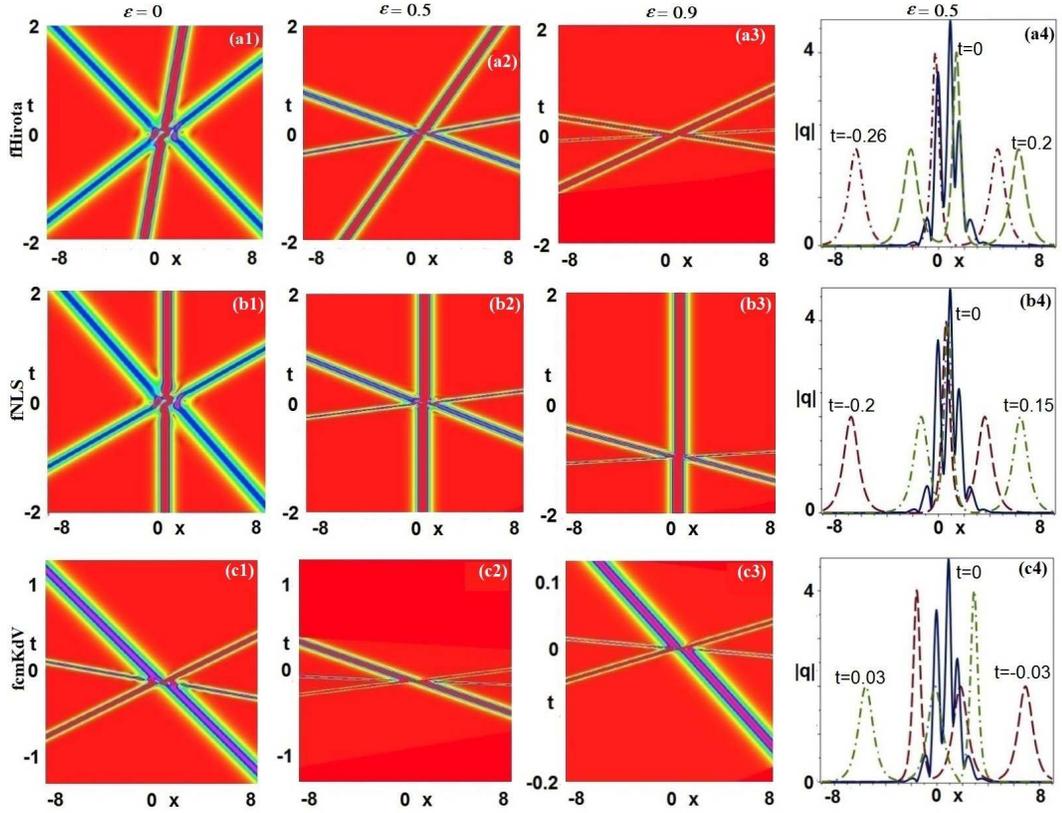}}}\hspace{-0.35in}
\vspace{0.15in}
\caption{Elastic interactions of fractional three-soliton solutions (\ref{recons}) with $N=3$ for $k_1=1+i,\, k_2=2i,\, k_3=-2+i, \, A_+[k_1]=A_+[k_2]=A_+[k_3]=1$ and $\epsilon=0,\, 0.5,\, 0.9$. (a1)-(a4) fHirota at $\alpha=1,\, \beta=0.05$; (b1)-(b4) fNLS at $\alpha=1,\, \beta=0$; (c1)-(c4) fcmKdV at $\alpha=0,\, \beta=1$.}
   \label{fig3}
\end{figure}

\v {\it Elastic interactions of fractional three-soliton solutions}.---For $N=3$, if we take three spectral parameters $k_1,\, k_2,\, k_3$ and $A_+[k_1]=a,\, A_+[k_2]=b,\, \, A_+[k_3]=c$, then the fractional three-soliton solutions of the fHirota equation are given by Eq.~(\ref{recons}).
As $\beta=0$, we have the fractional three-soliton  solution of the fNLS equation. When $\alpha=0$, we have the fractional three-soliton solution of the fcmKdV equation.

For the three taken spectral parameters $k_1=1+i,\, k_2=2i,\, k_3=-2+i$ and $A_+[k_1]=A_+[k_2]=A_+[k_3]=1$, Figs.~\ref{fig3}(a1-a4) display the fractional 3-soliton solution of the fHirota equation as $\alpha=1,\, \beta=0.05$, which imply the elastic interactions of two right-going
(left and center branches) and another left-going (right branch) travelling-wave solitons at $\epsilon=0,\, 0.5,\, 0.9$. Figs.~\ref{fig3}(b1-b4) display the fractional three-soliton solutions of the fNLS equation as $\alpha=0,\, \beta=1$, which also imply the elastic interactions of one left-going, one center-going  and another right-going travelling-wave solitons at $\epsilon=0,\, 0.5,\, 0.9$. Figs.~\ref{fig3}(c1-c4) show the fractional three-soliton solutions of the fcmKdV equation as $\alpha=0,\, \beta=1$, which are also elastic interactions of two left-going and another right-going travelling-wave solitons at $\epsilon=0,\, 0.5,\, 0.9$.

In fact, one can also study the other fractional $N\,(N>3)$-soliton solutions in terms of the formula (\ref{recons}).

\v {\it Fractional $N$-soliton solutions of fHONLS equations}.---Similarly, we can also solve the integrable fractional HONLS equations (\ref{fnlss}) by the IST with the matrix RH problem such that the fractional $N$-soliton solutions of Eq.~(\ref{fnlss}) is given by Eq.~(\ref{recons}), where $\theta_{\epsilon}$ is taken place by the general form
\bee\no
 \theta_{\epsilon}(x,t; k)=-kx+\frac12\left(\sum_{j=2}^N\alpha_ji^{\varsigma_j+j}(-2k)^j\right)|4k^2|^{\epsilon}t.
\ene

\section{Conclusions and discussions}

In conclusion, we have analyzed the explicit forms and anomalous dispersive relations of the fractional higher-order NLS euqations containing the fHirota, fcmKdV, and fLPD equations. We investigate the IST with the RH problem to study fractional $N$-soliton solutions of the fHirota and fcmKdV equations. Moreover, we analyze the elastic interactions of the fractional two- and three-soliton solutions. The wave, group, and phase velocities of these envelope fractional one-soliton solutions are shown to be related to the power law of their amplitudes. We also deduce the general formula for the fractional $N$-soliton solutios of integrable generalized fractional higher-order NLS equations. The used idea can also be extended to other integrable fractional nonlinear evolution equations. These results will be useful to display the super-dispersive transports of nonlinear waves in fractional nonlinear media. \\

\noindent {\bf Acknowledgments}

 We thank G. Zhang for useful discussions. The work was supported by the National Natural Science Foundation of China (No. 11925108).

\end{document}